# Efficacy of Statistical and Artificial Intelligence-based False Information Cyberattack Detection Models for Connected Vehicles

**Sakib Mahmud Khan, Ph.D.**
Assistant Research Professor
Glenn Department of Civil Engineering
Clemson University
200 Lowry Hall, Clemson, SC 29634
Email: sakibk@g.clemson.edu

**Gurcan Comert, Ph.D.**
Comp. Sci., Phy., and Engineering Department
Benedict College
1600 Harden Street
Columbia, SC 29204
Email: Gurcan.Comert@Benedict.edu

**Mashrur Chowdhury, Ph.D., P.E., F. ASCE**
Eugene Douglas Mays Endowed Chair in Transportation
Glenn Department of Civil Engineering
Clemson University
216 Lowry Hall, Clemson, South Carolina 29634
Email: mac@clemson.edu

**ABSTRACT**

Connected vehicles (CVs), because of the external connectivity with other CVs and connected infrastructure, are vulnerable to cyberattacks that can instantly compromise the safety of the vehicle itself and other connected vehicles and roadway infrastructure. One such cyberattack is the false information attack, where an external attacker injects inaccurate information into the connected vehicles and eventually can cause catastrophic consequences by compromising safety-critical applications like the forward collision warning. The occurrence and target of such attack events can be very dynamic, making real-time and near-real-time detection challenging. Change point models, can be used for real-time anomaly detection caused by the false information attack. In this paper, we have evaluated three change point-based statistical models; Expectation Maximization, Cumulative Summation, and Bayesian Online Change Point Algorithms for cyberattack detection in the CV data. Also, data-driven artificial intelligence (AI) models, which can be used to detect known and unknown underlying patterns in the dataset, have the potential of detecting a real-time anomaly in the CV data. We have used six AI models to detect false information attacks and compared the performance for detecting the attacks with our developed change point models. Our study shows that change points models performed better in real-time false information attack detection compared to the performance of the AI models. Change point models having the advantage of no training requirements can be a feasible and computationally efficient alternative to AI models for false information attack detection in connected vehicles.

**Keywords:** cybersecurity, connected vehicle, statistical model, AI, change point models

## INTRODUCTION

Connected vehicles are a part of the transportation cyber-physical systems (TCPS). As shown in Figure 1, in a connected transportation cyber-physical system (TCPS), mobile edges (such as connected vehicles or CVs, connected transportation system users), fixed edge devices (such as roadside unit), and backend servers (such as cloud/on-site servers) are integrated via different communication options (*1*). Heterogeneous data are exchanged between these edge devices while running multiple real-time TCPS applications. While the seamless interconnection holds promises of safety, operational, environmental, and energy efficiency benefits, the external connectivity also increases the exposure to cyberattacks. These cyberattacks include jamming, spoofing, denial-of-service (DoS), malware injection, blackhole, eavesdropping, Sybil attacks, and false information. The dynamic behavior of these cyberattacks makes real-time threat detection challenging. Attacks on TCPS applications can have a significant impact on public safety. For example, once false information is injected into a connected vehicle, a critical safety application, such as the collision warning, can malfunction that can cause catastrophic consequences. To detect these anomalies in real-time, both change point models (*2*) and data-driven artificial intelligence (AI)-based models can be used (*3*). The focus of our research is to study and compare the performance of the statistical change point models and data-driven AI models for detecting false information cyberattacks on CV in a TCPS environment.

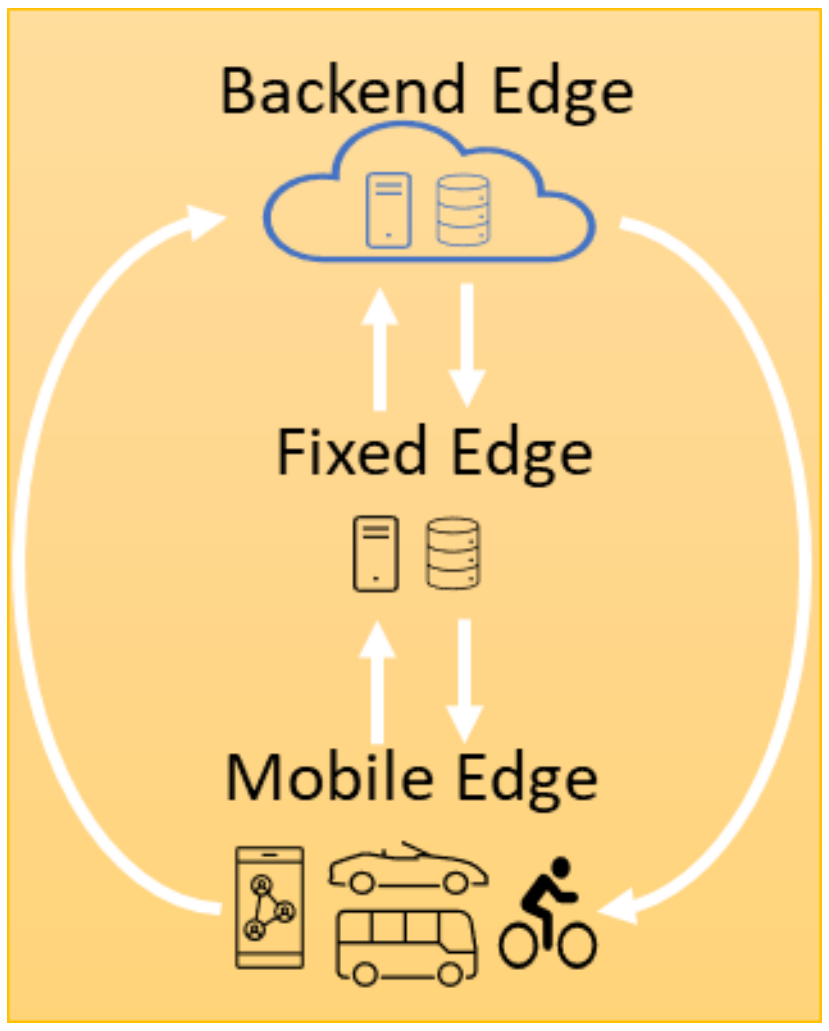

**Figure 1 TCPS layers in a connected environment**

One way to detect false information cyberattacks in the continuously streamed aggregated data in connected vehicles is to track the breakpoints in the time series data where the breakpoints represent discrete shifts caused by dynamic changes in the data series. These switching inside the data flow can be captured in a model which has a parameter based on the fluctuating pattern. For example, according to the discussion in (*4*), a structural break or change occurs in the model of the form $y_t = \mu + \phi y_{t-1} + \varepsilon_t$, if one (or more) of the parameters (i.e., constant $\mu$, $\phi$, or white noise's $(\varepsilon_t)$ mean or variance) changes (*5*). While considering traffic flow events, it should be reiterated that such dynamic and discrete shifts are common in traffic flow considering different traffic events. Such events include the occurrence of a sudden incident or suddenly arriving at a work zone and a sudden rise in demand. Consequently, with this description of the data generation process, model parameters need to be updated in relation to the breakpoints. One efficient way to identify the changes in the time series data is to deploy a statistical detection model where parameters will be continuously tuned based on the changes in the streaming data. One approach is of parameter update is to implement the rolling horizon approach, where changes are constantly monitored within a rolling and predefined time window. However, such approach will not be able to detect real-time and sudden fluctuations in the data if the time-window is relatively large, as considered in (*6*). Also within a temporal time window, the rolling horizon approach fuses observations from different regimes (i.e., attack and no-attack) and generates weighted average estimates based on the observations from different regimes. Combining data from multiple regimes within a large temporal window compromises the reliability of the cyberattack detection (*7*). In this context, change-point detection/clustering algorithms, such as the Expected Maximization or EM, Cumulative Sum or CUSUM algorithm (*8*), Hidden Markov Models and Bayesian Online Change Point Detection (BOCPD) (*9*), are studied to detect shifts in the cyberattack detection process in real-time. Another cyberattack detection process used the data-driven and supervised AI models, which are popular classification models to identify the attack data based on the underlying data characteristics (*10*). In a supervised learning model, the input feature $x$ and output label $y$ are known for certain circumstances. Let us assume that we have $N$ training examples. $N$ consists of the datapoints $\{(x_1, y_1), (x_2, y_2), \ldots, (x_N, y_N)\}$ where $x_i$ is the feature (can be a feature vector) and $y_i$ is the label of the $i^{th}$ sample.

Using the dataset with $N$ data points, a supervised learning model will develop a function $f: X \rightarrow Y$ to correctly classify the data points where $X$ is the input space with $x$ features and $Y$ being the output space with $y$ labels.

Our research objectives are to (i) develop an efficient and reliable cyberattack detection framework for false information detection for connected vehicles using change point and AI-based models and (ii) evaluate the performance of the models for false information detection. Our edge-centric cyberattack detection module will be in an infrastructure edge, which is receiving broadcasted Basic Safety Messages from connected vehicles according to the SAE 2745 standards (*11*). This study focuses on a V2X application called the vehicle-based traffic surveillance application, as it is the foundation of many connected vehicle applications (*12*). This application can be considered as a foundational CV application as numerous other CV applications can leverage data collected through the vehicle-based surveillance application considered in this study. Here all CVs share basic safety messages or BSM with the roadside devices where the data are aggregated for each timestamp. Figure 2 shows a scenario where the attacker has gained access to a targeted CV to manipulate the BSM data of the CV. After aggregating the BSM data (such as speed for each timestamp), the roadside unit will forward them to the backend server, which can be considered as a cloud server. Utilizing this aggregated data, if a CV safety application generates wrong output, it could lead to wrong outcomes like inaccurate incident detection, false collision warning, and undesirable traffic routing. A cyberattack detection module can be implemented in any infrastructure edge that has computing resources and communication capabilities. These modules can be used to identify anomalies in the aggregated BSM collected from connected vehicles (*13*). With the purpose of developing security controls for connected vehicles that afford the right level of protection given the data and means of transmission, we have developed and evaluate the false information attack detection models in this research.

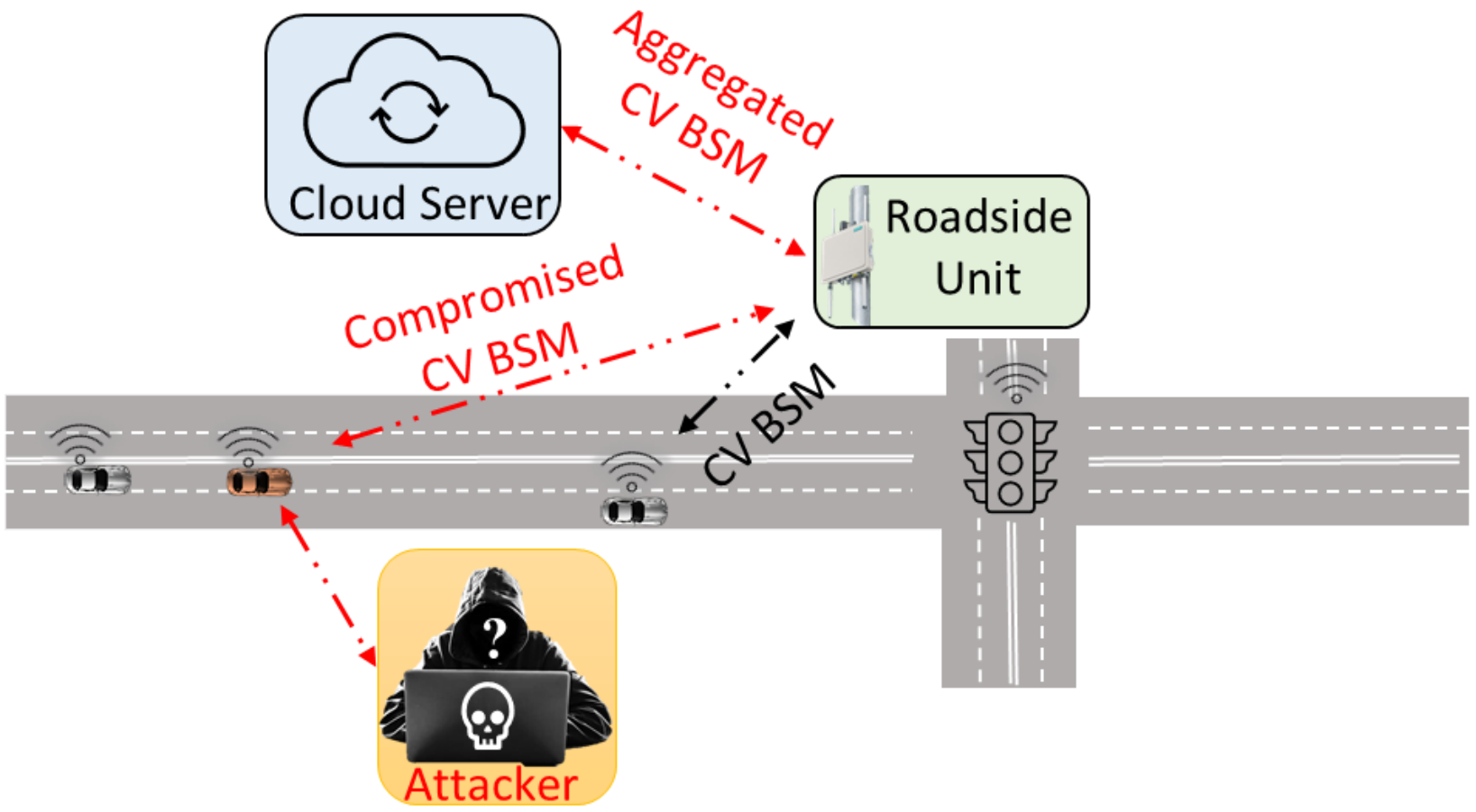

**Figure 2 Information flow in a compromised CV environment (BSM: Basic Safety Message)**

Earlier research on this topic did not evaluate and compare statistical and AI-based cyberattack detection models focusing on CV applications. While addressing this knowledge gap, we have considered the cyberattack detection model to run in the infrastructure edge device to secure CV applications against false information attacks. The following sections discuss related studies, attack detection models, the evaluation scenario, and findings from this research.

## RELATED WORKS

### Change Point Models for Cyberattack Detection

With more data getting available, one of the inferential problems is to detect changes in the distributional properties of the incoming data. In this context, we have seen earlier applications of detecting changes in the volatility (i.e., a statistical measure of dispersion) of time series in genomics (*14*) and finance (*15*). There is a growing need to identify these change point locations, characteristics before and after the change points over time and search for such efficient algorithms (*16*, *17*). This retrospective analysis consists of segmentation of the data and identification of the unknown number of change points. Change point has been in the literature since 1974 (*18*, *19*). There exist many algorithms to determine the number of change points with minimum computational time requirements. In the first step of our research, offline analysis of the data will be carried out to determine the distributional properties such as type of changes, the differences among the mean levels, and the ratio of standard deviations at a particular regime. Changes in the generative parameters are often key aspects of time series data, which comprise many distinct regimes. An inability to react to regime changes can have a worsening impact on forecasting performance. Change point detection attempts to reduce this impact by adopting the model parameters to the changes and thus can detect CV cyberattacks (*2*).

## Artificial Intelligence Models for Cyberattack Detection

Recent revolutions in computational hardware technologies and distributed computation facilities (e.g., cloud-edge collaboration) have led to numerous AI-based cybersecurity innovations that can be used to protect the connected in-vehicle systems and associated communication systems. AI models have been used for identifying cyberattacks on cyber-physical systems in earlier studies (*20*–*22*). For intrusion detection in a CV environment, neural networks are used (*20, 23*). Recurrent neural network models are also used for anomaly detection in in-vehicle networks (*24*). Here, the authors explored how neural networks can be used to improve the security of the controller area network (CAN) for vehicles. Six state-of-the-art AI classification models are used in the research as these models, once trained, are readily implementable for online false information cyberattack detection and can benchmark the expected cyberattack detection performance of the change point models.

# METHOD

## CV Cyberattack Type

For this research, we have used the false information attack to create anomalies in the CV Basic Safety Message (BSM) data. The BSM part 1 data are broadcasted ten times per second (*25*) and include different data items, such as CV position, heading, speed, and heading. In a false data injection or false information attack scenario, an attacker inserts malicious information that contradicts real information (*26*). In this study, we model the attack in such a way where an attacker after getting access to a CV, inserts false velocity data in the attacked CV BSM, as shown in Figure 3. The attacker's intent is to disrupt traffic or even triggering a collision. In our study, we have generated compromised speed data from a connected vehicle under attack and broadcasted them at 10 Hz or 10 packets/sec to an infrastructure edge which is the roadside unit or RSU. RSU aggregates the data by averaging the data per 0.1 seconds. The attack detection models can run in any infrastructure edge (RSU) to identify if an attack has occurred on one of the CV BSM data (speed in this study). Other BSM data, such as location, acceleration are not affected by the attack in our study.

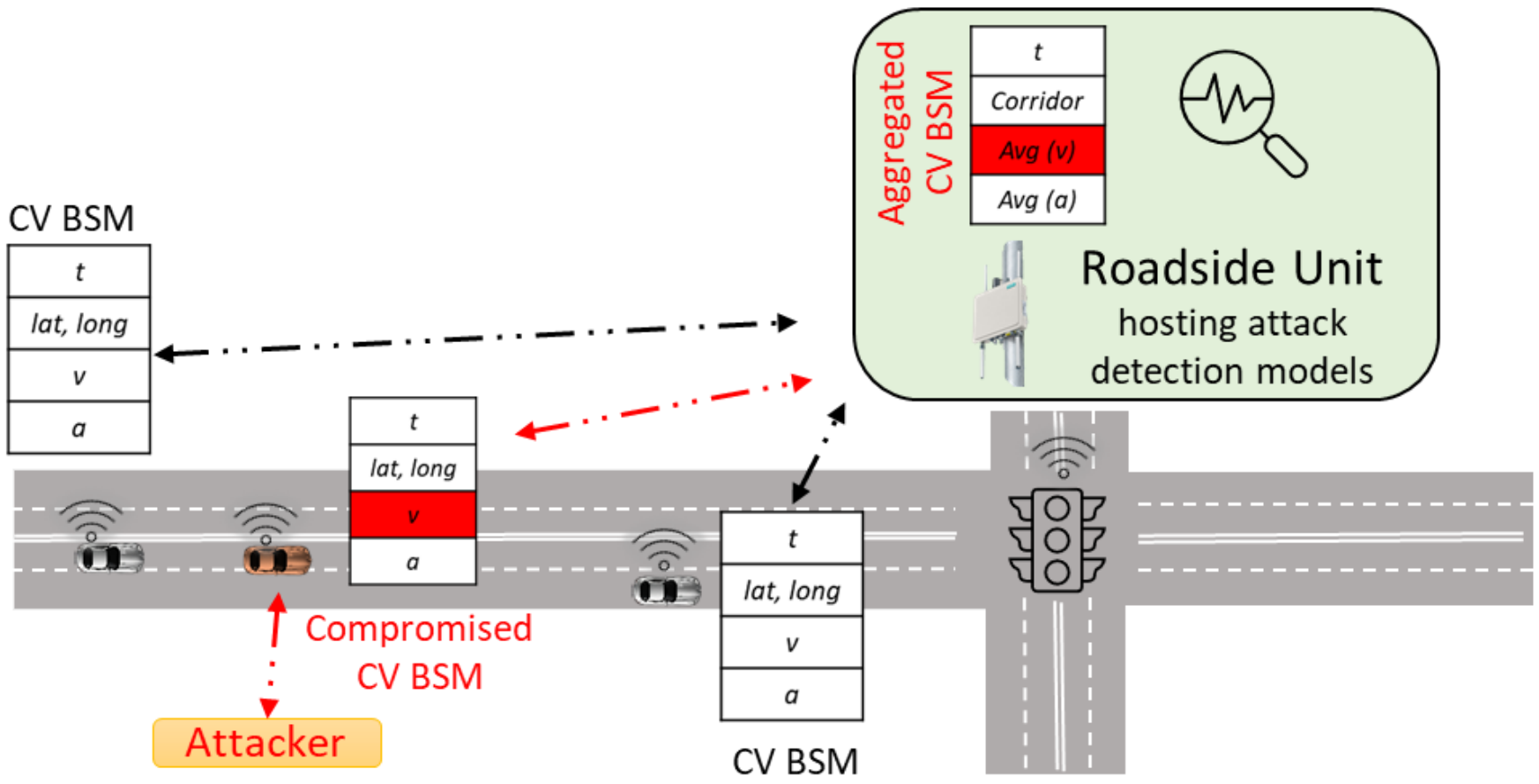

**Figure 3 False Information Attack on CV BSM**
**(t: timestamp, v: speed, a: acceleration)**

## Cyberattack Detection Models

### *Change Point Models*

Our goal is to develop resilient attack detection methods using classical hypothesis testing-based change point detection algorithms and the Bayesian reliability-based method, which is Bayesian Online Change Point Algorithm (*9*). We first need to identify efficient and effective cumulative measures to track any change in the data. Thus, we will determine what critical information needs to be tracked in a standard BSM (*27*). The change-point models are developed and evaluated in R.

Bayesian Online Change Point Detection (BOCPD)

A BOCPD algorithm was introduced by Adams and Mackay (*9*). It is given in Algorithm 1. The hazard function or hazard rate *h(i)* (as simple as a constant $\lambda$, e.g., *0.01*) describes how likely a change point is given the current run length *r*. For a lifetime distribution $f(i) = \lambda e^{-\lambda i}$ and survival function $s(i) = P(T > i) = 1 - \int_0^i f(t)dt = e^{-\lambda i}$, hazard function is defined as $f(i)/s(i)=\lambda$. Mean lifetime is $1/\lambda$, which is considered to be *100* time intervals. This approach is a completely different logic than other change point algorithms used in transportation literature. Essentially, the hyperparameters are representing the underlying data generating processes' $\Theta=\{\mu_0, \kappa, \alpha, \beta\}$, and they are updated as we get new observations and help us to calculate the probability of change points given run length and observation. We slightly revised the algorithm to be able to use real-time for attack detection in CVs. The classic BOCPD algorithm includes an optimization step to find "0" probability that is change point evidence. This requires large memory as it is generating and searching as large as the $N \times N$ matrix. Instead, we calculate a single probability for the obtained observation and provide a reasonable threshold (e.g., *p(attack observation)<0.0002*). Certainly, this approach adds another parameter (i.e., threshold value) to be tuned. In fact, the algorithm becomes very close to the way we utilize EM sensitive to initial values while it runs considerably faster.

In Algorithm 1, we assume Normal-Inverse Gamma prior ($\mu \sim N\left(\mu_0, \frac{\sigma^2}{\kappa}\right), 1/\sigma^2 \sim Ga(\alpha, \beta)$ ) with hyperparameters $\{\mu_0, \kappa, \alpha, \beta\}$ and derivations are resulting predictive *t*-distribution given previous observations ($X_{1:i}$) and run length at time step $i$ ($r_i$) ($p(Y_{i+1}|Y_{1:i}, r_i) \sim t_{2\alpha}(\mu, \frac{\beta(\kappa+1)}{\alpha\kappa}$) degrees of freedom $2\alpha$, mean $\mu$ and variance $\frac{\beta(\kappa+1)}{\alpha\kappa}$) with online Bayesian parameter updates (*28*). Predictive distribution can be used as Gaussian processes and EM responsibilities.

**Algorithm 1 BOCPD for detection in series of *N* observations (*9, 27–29*)**

Initialize parameters, $\Theta = \{\lambda=0.01, \mu_0=0, \kappa = 0.1, \alpha = 10^{-5}, \beta = 10^{-5}\}$
 Define $H$, function
 **for** $X_i \ni i$ in $1{:}N$ **do**
  Calculate prior: $\pi_i^r = p(Y_i|Y_{1:i}, r_i) \sim t_{2\alpha}(\mu, \frac{\beta(\kappa+1)}{\alpha\kappa})$
  Calculate non-CP: $p(r_i = r_{i-1} + 1, Y_{1:i}) = p(r_{i-1}, Y_{1:i-1})\pi_i^r(1-h)$
  Calculate CP: $p(r_i = 0, Y_{1:i}) = \sum_i p(r_{i-1}, Y_{1:i-1})\, \pi_i^r h$
  Normalize: $p(r_i = 0, Y_{1:i}) = p(r_i = 0, Y_{1:i}) / \sum_i p(r_{i-1}, Y_{1:i-1})$
  Find Maximum CP or low posterior: $\text{argmax}_{p(r_i=0, Y_{1:i})}$
  **if** $CP_{i+1}^+ < CP_i^+$ **then**
   $Y_{i+1}$ is a CP/an Attack, $i' = i$
   Set run length $r_{i+1} = 1$
   $\bar{Y}_i = Y_i$
  **else** Updates

```
        r_{i+1} = r_i + 1
        Ȳ_i = (1 − p(r_i))Ȳ_{i':i} + p(r_i)Y_i
        μ = p(r_i)Y_i + (Ȳ_i + r_i μ)(1 − p(r_i))/(r_i + 1)
        κ = κ + 1
        α = α + 1/2
        β = β + κ(Y_i − Ȳ_i)^2/(2(κ + 1))
    end if
    p(Y_{i+1}|Y_{1:i}, r_i) ~ t_{2α}(μ, β(κ+1)/(ακ)) evaluated at X_i with updated {μ,κ,α,β}
end for
```

Expectation Maximization (EM)

EM was originally used for parameter estimation when point estimations are not tractable. It can provide maximum likelihood estimates of complex distributions' parameters such as finite mixtures. However, for false information attack detection, we use EM for classification and use its responsibilities to obtain the probability of attack or no attack events. To achieve, first we generate some sample data that will have examples of normal and abnormal observations. In order, reduce dependency to initial parameters, we have used the first ten (*N-1*) values of the data to generate seven no attack data ($X_{1:7} \sim N(\bar{Y}_{1:10}, S^2_{Y_{1:10}})$) and three significantly different values to represent possible unknown attacks, for instance $X_{8:10} \sim N(0.5, 1^2)$. So, we generated a vector of values size *N-1* as $X_{1:10} = [X_{1:7} \sim N(\bar{Y}_{1:10}, S^2_{Y_{1:10}}), X_{8:10} \sim N(0.5, 1^2)]$ and $N^{\text{th}}$ ($11^{th}$) value is the new observation $Y_i$. We can initialize using small no attack sample data ($X_{1:10}$) or by the expert opinion of no attack, attack values. Thus, we initialize parameters for: (i) Gaussian no attack observations, and (ii) attack mean and variance, and proportion are denoted as $\Theta = (\mu_1, \mu_2, \sigma_1, \sigma_2, \pi)$. Then, until convergence, we calculate: responsibilities *for N=1,..,11* where last value $X_{11}$ is $Y_i$, and we calculate the probability of attack of $p(Y_i \sim N(\hat{\mu}_2, \hat{\sigma}_2^2)$ as we better estimate parameters of mixture of 2 Normal distributions.

**Algorithm 2 EM revised for detection in series of *N* observations (*31*)**

Get two classes of observations $X_{1:11} = [X_{1:7} \sim N(\bar{Y}_{1:10}, S^2_{Y_{1:10}}), X_{8:10} \sim N(0.5, 1^2), Y_i]$

Initialize parameters, $\Theta = (\mu_1, \mu_2, \sigma_1, \sigma_2, \pi)$ *(alternatively from* $X_{1:10}$*)*

**for** $Y_i \ni i$ in $1{:}N$ **do**

  **for** *j* in $1{:}11$ **do**

$$p(Y_i \sim N(\hat{\mu}_2, \hat{\sigma}_2^2) = \hat{\pi}\phi(Y_i|\hat{\mu}_2, \hat{\sigma}_2^2)/[\hat{\pi}\phi(Y_i|\hat{\mu}_2, \hat{\sigma}_2^2) + (1-\hat{\pi})\phi(Y_i|\hat{\mu}_1, \hat{\sigma}_1^2)]$$

  **end for**

$$\hat{\mu}_1 = \frac{\sum_{i=1}^{N}(1-p(Y_i \sim N(\hat{\mu}_2, \hat{\sigma}_2^2)Y_i)}{\sum_{i=1}^{N}(1-p(Y_i \sim N(\hat{\mu}_2, \hat{\sigma}_2^2))} \quad \hat{\mu}_2 = \frac{\sum_{i=1}^{N} p(Y_i \sim N(\hat{\mu}_2, \hat{\sigma}_2^2)Y_i}{\sum_{i=1}^{N} p(Y_i \sim N(\hat{\mu}_2, \hat{\sigma}_2^2)}$$

$$\hat{\sigma}_1^2 = \frac{\sum_{i=1}^{N}(1-p(Y_i \sim N(\hat{\mu}_2, \hat{\sigma}_2^2)(Y_i-\hat{\mu}_1)^2)}{\sum_{i=1}^{N}(1-p(Y_i \sim N(\hat{\mu}_2, \hat{\sigma}_2^2))} \quad \hat{\sigma}_2^2 = \frac{\sum_{i=1}^{N}(p(Y_i \sim N(\hat{\mu}_2, \hat{\sigma}_2^2)(Y_i-\hat{\mu}_2)^2)}{\sum_{i=1}^{N}(p(Y_i \sim N(\hat{\mu}_2, \hat{\sigma}_2^2))}$$

$$\hat{\pi} = \sum_{i=1}^{N} p(Y_i \sim N(\hat{\mu}_2, \hat{\sigma}_2^2)Y_i/N$$

  **if** $p(Y_i \sim N(\hat{\mu}_2, \hat{\sigma}_2^2) > 0.01$ **then** $Y_i$ is an attack

**end for**

CUSUM Algorithm

The direct usefulness of CUSUM algorithm lies in the quality control where the mean process level is monitored by the CUSUM chart or algorithm. In this study, CUSUM is used to monitor the expected level of the observed measure while detecting changes, thus, cyberattacks. Let's assume an identical-independently-distributed variable $Y_i$ having known $(\mu_1, \sigma^2)$ and $\mu_2$, a new process mean of the variable $Y_i$ will be estimated after observing a possible shift. We have selected CUSUM parameters based on the findings from literature: $K = \delta\sigma/2$, $H = 5\sigma$ and $\delta = 1$.

Based on (*32*), we have selected the adaptive version of CUSUM. The adaptive model is used for processes having other than zero mean, and it has a single weight parameter ($\alpha$). In Algorithm 3, $D_i = (\bar{\mu}_i - \mu_1)$ and $\bar{\mu}_i = \alpha\bar{\mu}_{i-1} + (1-\alpha)Y_i$ at time step $i$. For less false positive detection, we have considered $H = 5\sigma$. The $C_i^+$ and $C_i^-$ represents the positive deviations (values above the target), and negative deviations (values below the target), respectively. In this study, we do not assume normal mean and standard deviations as known. We calculate them from a few normal values as $\mu = \bar{Y}_{1:3}$, $\sigma = \sqrt{S^2_{Y_{1:3}}}$.

**Algorithm 3 CUSUM revised for detection in series of *N* observations (*32*)**

1: Choose parameters, $\mu = \bar{Y}_{1:3}$, $\sigma = \sqrt{S^2_{Y_{1:3}}}$, $K = \delta\sigma/2$, $\delta=1$, $\alpha=0.025$

2: Choose parameters, $H$ is set to $5\sigma$ *(33)*

3: **for** $Y_i \ni i$ in $1{:}N$ **do**

4: $$C_i^+ = \left[0, C_{i-1}^+ + \frac{\alpha D_i}{\sigma^2}[Y_i - D_i - \alpha D_i/2]\right]^+$$

$$C_i^- = \left[0, C_{i-1}^- - \frac{\alpha D_i}{\sigma^2}[Y_i + D_i + \alpha D_i/2]\right]^+$$

$$D_i = \bar{\mu}_{i-1} - \mu_i$$

$$\bar{\mu}_i = \alpha\bar{\mu}_{i-1} - (1-\alpha)Y_i$$

5: **if** $C_i^+ > H \vee C_i^- > H$ **then** $Y_i$ is an attack, set: $C_i^+ = C_i^- = 0$

6: **end for**

Note that attack and no attack data should ideally have two separate (far away modes or peaks) distributions. However, in reality, we can find almost identical (i.e., very close modes) mixtures of distribution. To separate these, we need a transformation with the help of covariates. In fact, this is the same idea as identifying key features in classification problems. In this paper, since detection is in real-time, by the rolling transformation of the variable, new $Y_i$ can be obtained as $Y_i = Var(\Delta[\overline{\Delta S}_{i-9:i} - 0.99(\Delta A_i - \overline{\Delta A}_{i-9:i})])$, where $\Delta$ is the difference, *Var(.)* variance of the sample data, $A_{i-9:i}$ acceleration values for our case study between time *i-9 to i* values, $\overline{\Delta S}_{i-9:i}$ sample mean of the $S_{i-9:i}$ speed values (which is under attack) for our case study. Note that $\overline{\Delta S}_{i-9:i} - 0.99(\Delta A_i - \overline{\Delta A}_{i-9:i})$ is the control variate method commonly used for variance reduction in simulations *(34)*. We aimed to reduce variation with and without attack data sections.

*Artificial Intelligence Models*

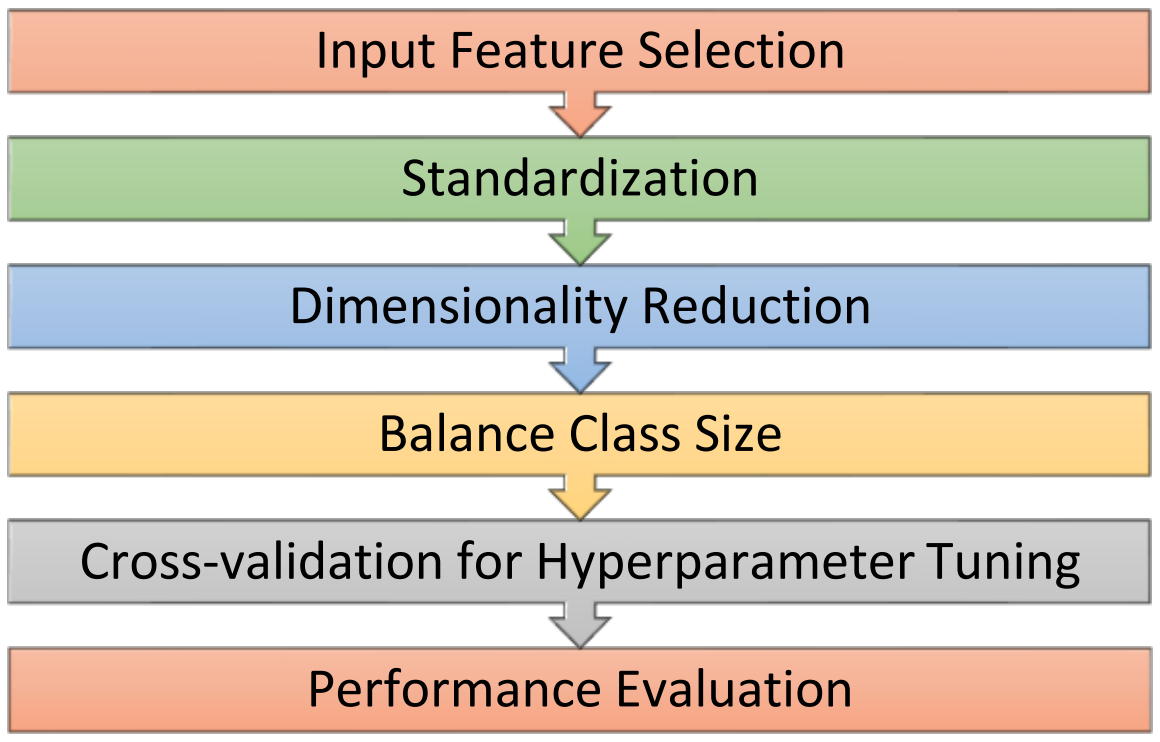

**Figure 4 AI model development and evaluation pipeline**

The data-driven machine learning models rely on the underlying data characteristics to classify the false information attack and no-attack data. BSM data contains different types of data with different distributions, and attacks can happen in any arbitrary order on any of the BSM data types (e.g., speed, acceleration, etc.). To properly classify the data with machine learning models, we need to do proper data processing and feature selection. We also need to take care of the class imbalance issue so the AI models do not show bias when detecting cyberattacks on CVs. In the analysis step, we have created a pipeline to train and test the AI models, as shown in Figure 4. We have considered two features (i.e., average speed and average acceleration) to detect false information attacks on connected vehicles. At first, each feature is standardized separately, and the calculated standardized value for a sample *x* is $\frac{x-\mu}{s}$ with $\mu$ being the mean of the sample feature and *s* is the standard deviation of the feature. Later the dimensionality of the features is reduced using Singular Value Decomposition of the data. For the class size balancing, we have used two methods. For support vector machine, decision tree, and random forest classifier, the class weights during hyperparameter tuning are automatically adjusted so the weights become inversely proportional to class frequencies. For K-nearest neighbor, Extreme Gradient Boosting, and neural network, we have used Synthetic Minority Over-sampling Technique or SMOTE-integrated class size balancing methods, which balances by a combined approach of under-sampling of the majority class and over-sampling the minority class *(35–37)*. These balancing methods are selected based on multiple trials. We have used 5-fold cross-validation techniques with an *80%-20%* split of training and test data for hyperparameter tuning. Using the grid search method, the hyperparameters are chosen when they provided the highest accuracy *(38)*. We have used Keras *(39)* to develop the neural network model, the XGboost to develop the extreme gradient boosting model *(40)*, and the Scikit Learn *(41)* to develop the other models. The pipeline in Figure 4 is developed with Scikit learn.

<u>K-nearest Neighbors (KNN)</u>

The supervised KNN classifier detects false information attacks based on the nearest neighbors of each query point with specific average speed *(v)* and average acceleration *(a)* values. Let's assume that data point $x_t(v_t, a_t)$ is the test data for a particular timestamp *t*. At time *t*, an attack will be detected if the *k* nearest training samples have a majority of attack labels.

<u>Support Vector Machine (SVM)</u>

SVM classifier will detect the false information attack by learning to create a hyperplane to differentiate the two classes (i.e., false information attack and no-attack), based on the training data. The following is the equation of the hyperplane for binary classes (class +1 or attack and class -1 or no-attack):

$$\boldsymbol{x_i}.\boldsymbol{w} + b \geq +1, if\ y_i = +1 \quad (1)$$

$$\boldsymbol{x_i}.\boldsymbol{w} + b \leq -1, if\ y_i = -1 \quad (2)$$

Here ***w*** is the weight vector, ***x*** is the input vector with average speed *(v)* and average acceleration *(a)*, *i* is the index of the input feature, *b* is the bias, and *y* is the output. Maximizing the boundary width of the binary classification hyperplane eventually becomes a constrained optimization problem that can be solved by the Lagrangian Multiplier method. The following equation shows the maximization problem of the Lagrangian Multiplier *L* with the dual variable, $\alpha$, and input datapoints *i*, *j*.

$$Max\ L = \sum \alpha_i - \frac{1}{2}\sum \alpha_i \alpha_j K(\boldsymbol{x_i}, \boldsymbol{x_j}) \quad (3)$$

Here $K(\boldsymbol{x_i}, \boldsymbol{x_j})$ is the Kernel function which implicitly transforms datasets into a higher-dimensional space utilizing minimum computational resources. The selection of the appropriate Kernel function is conducted based on cross-validation.

Neural Network (NN)

We have developed a feed-forward neural network model that consists of three layers, including input, hidden and output layers. The input layer receives the data, the hidden layer does the computation, while the output layer generates the attack classification output. During the training phase, the data goes in the forward direction from the input layer to the output layer and using back propagation, the model parameters (weight and bias vectors) are optimized. The computations taking place at every neuron in the output *(o)* and hidden *(h)* layer are as follows,

$$\boldsymbol{h_x} = f_h\ (\boldsymbol{w_h}.\boldsymbol{x}+\boldsymbol{b_h}) \quad (4)$$
$$\boldsymbol{y} = f_o\ (\boldsymbol{w_o}.\boldsymbol{h_x}+\boldsymbol{b_o}) \quad (5)$$

$\boldsymbol{x}$ is the N-dimensional input vector with average speed *(v)* and average acceleration *(a)*, and $\boldsymbol{y}$ is the output vector. $\boldsymbol{w_h}$ is the weight vector. $\boldsymbol{b}$ is the bias vector. *f* is the non-linear activation function

Decision Tree (DT)

We have used the Classification and Regression Trees (CART) algorithm *(42)* as the decision tree classifier to detect the false information cyberattack on CV BSM. Using this model, the input variable X is iteratively split into homogeneous intervals based on the output label Y. The split is done to maximize the information gain that measures how well the attack classes are split. Impurity function is used while computing the information gain, and one widely accepted impurity function is the Gini Impurity Index.

Let's assume for our binary classification, a node *t* is classified into two classes which are $c_1$ (attack) and $c_2$ (no-attack). The Gini Impurity Index or $G_{c_1}$ is estimated using the following equation:

$$G_{c_1}=\sum_{c_1,c_2} \varepsilon(c_1|c_2)\ .\ p(c_1|t).\ p(c_2|t) \quad (6)$$

Here $\varepsilon(c_1|c_2)$ is the misclassification cost when a $c_2$ class case is misclassified as a $c_1$ class case. $p(c_1|t)$ is the probability of a case in class $c_1$ given the case falls into the *t* node. With any specific splitting scenario and a specific node, the goal is to decrease the Gini impurity as below:

$$\Delta G=G_{c_1}\text{-}\ p_L G_{c_1,L}\text{-}p_R G_{c_1,R} \quad (7)$$

Here $p_L$ is the probability of sending the case to the left child node $L$ and $G_{c_1,L}$ is Gini impurity Measure for left child node. Once one node is split, the same splitting process is done with its child node. When no information gain is possible, the decision tree classifier stops the splitting.

Random Forest (RF)

Considering multiple decision trees to detect the false information attack, random forest (an ensemble classifier) reduces the variance and possibility of overfitting caused by a single decision tree model *(43)*. Using the bagging or bootstrapping sampling technique, random samples (with replacement data point meaning same data point can be used multiple times) are drawn to create multiple decision tree models. The final classification is done based by averaging the probabilistic prediction of individual decision tree classifiers.

Extreme Gradient Boosting (XGB)

Like the RF, XGB *(44)* is an ensemble learning model. The XGB conducts sequential learning by fitting each tree in the sequence to the earlier tree's residuals. While fitting the new tree, XGB uses a gradient descent algorithm to minimize the loss. The XGB includes regularization factors that help to reduce the possibility of overfitting.

### Case Study Area

For this research, we have used the same case study area as discussed in (*2*), which presented the calibration process of the SUMO simulated corridor. The test area in the South Carolina-Connected Vehicle Testbed (SC-CVT), a 2.34 mile urban arterial with multiple signalized intersections where CVs communicate with the roadside units or RSUs using DSRC (*11*). The testbed has 35 miles per hour speed limit. In the simulation, we have 200 vehicles per hour per lane based on the data collected from SC-CVT for the simulation period considered in this study. As SUMO cannot simulate DSRC communication between the vehicles and infrastructure edge for vehicle-to-infrastructure communications, we introduced latency in the CV-to-infrastructure edge BSM delivery time to simulate the DSRC communication from our earlier filed study presented in *(45)*.

## ANALYSIS AND FINDINGS

Here we discuss the findings from our false information cyberattack detection analysis to evaluate the performances of two types attack detection models: change point and AI models

### Model Parameter

For the change-points models, we have considered changing the initial parameters and studied the impact as detection algorithms prone to normal conditions input. Table 1 summarizes our findings that result best for the false speed cyberattack. We can see that BOCPD is able to detect the false information attack when the only normal mean level of the time series data are added into the model.

**Table 1 Change Point Model Parameters for False Information Detection**

| EM | BOCPD | CUSUM |
|---|---|---|
| $\theta_1 = (\bar{Y}_{1:10}, S^2_{1:10})$, $\theta_2 = (0.5, 1^2)$, $\pi = 0.80$<br>$X_{1:7} \sim N(\bar{Y}_{1:10}, S^2_{1:10})$, $X_{8:10} \sim N(0.5, 1^2)$ | $\lambda$=0.01, $\mu_0$ = 0.0,<br>$\kappa$ = 0.1<br>$\alpha=10^{-5}$, $\beta=10^{-5}$ | $\mu_1 = \bar{Y}_{1:3}$<br>$\sigma = \sqrt{S^2_{Y_{1:3}}}$ |

In our AI model analysis, we have a total of 2000 data points to classify the attack and no-attack scenarios. After 80-20 split in training and test data, we conducted a 5-fold grid search to find the optimized parameters and hypermeters, which are shown in Table 2.

**Table 2 Optimized AI Model Hyperparameters for False Information Attack Detection**

| AI Models | Optimized Model Parameters and Hyperparameters |
|---|---|
| **KNN** | Classification algorithm: ball tree, distance calculation method: Euclidean, # of neighbors: 19 |
| **SVM** | Cost coefficient C: 1024, kernel: radial basis, gamma: 2.45 |
| **NN** | Activation function to get output from each layer input: 'relu', the learning rate of each iteration to reduce cost function: 0.2, Data dropping inside the layers: 0.0, # of neurons in hidden layer: 10, initialization of random weight assignment: 'uniform,' optimizer to tune the hyperparameter: 'Adam', loss function: 'binary cross-entropy,' # of hidden layer: 1, epoch or # of times NN will work through the entire training dataset: 100, batch size or # of samples to process before updating the internal model parameters.: 50 |
| **DT** | Split criteria: entropy, max depth of the decision tree: 8 |
| **RF** | Bootstrap: True, max depth of the decision tree: 90, min # of samples for a leaf node: 5, minimum # of samples required to split an internal node: 12, # of trees in the forest: 400 |
| **XGB** | Classifier Objective: Logistic regression for binary classification, # of gradient boosted trees: 300, the learning rate of each iteration to reduce cost function: 0.3, max depth of the tree: 10, subsampling ratio of column while constructing each tree: 0.11, tree construction algorithm: exact, booster method: gbtree, Min loss reduction required to further partition a leaf node of the tree: 0.05 |

**Performance of Cyberattack Detection Models**

*Computational Performance*

Table 3 shows the summary findings of the computational time required to train the AI models. The total training time depends on the overall underlying data complexity and the search space of the parameters and hyperparameter sets. DT models take less time to train, as well as the KNN and SVM models. We have more parameters for XGB than RF, so the total training time for XGB is higher than RF, however as XGB is a computationally efficient algorithm, the average training time is much less than RF. NN models consumed the most time to train among all the models.

**Table 3 Summary of computation time in seconds for AI model training**

| **AI Models** | **Computation time for training (s)** | | |
|---|---|---|---|
| | Total time | Average time to process single-fold validation | St. Dev for single-fold validation processing |
| **KNN** | 33.0 | 0.21 | >0.001 |
| **SVM** | 44.4 | 0.17 | 0.40 |
| **NN** | 28206.0 | 3.26 | 302.98 |
| **DT** | 2.0 | >0.001 | >0.001 |
| **RF** | 824.0 | 1.14 | 0.99 |
| **XGB** | 7880.0 | 0.15 | 0.16 |

However, for inference, all models can efficiently classify the test data in real-time without any delay. Figure 5 shows the inference time of all models in milliseconds. EM model needs the highest time to classify the data. Overall the inference time for the change point model is a little higher than those of the AI models, however both can be used for real-time false information attack detection.

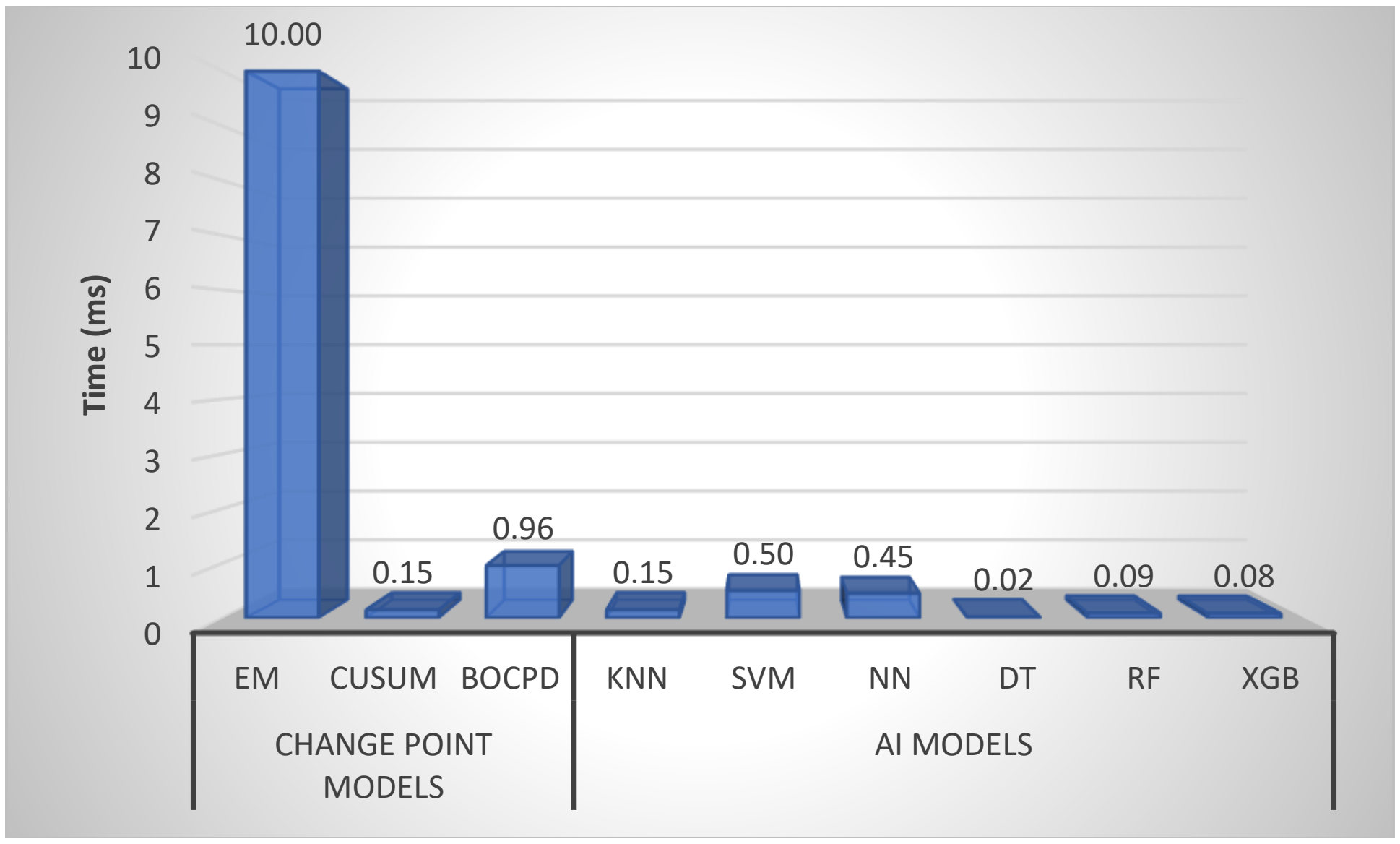

**Figure 5 False information detection time per new aggregated datapoint**

*Attack Detection Performance*

We have set $P(X_{t+1}|X_t, r_t)<0.0002$ as a change or attack for BOCPD and 0.01 for EM. CUSUM's threshold was set to $H= 5\sigma=5\sqrt{S^2_{Y_{1:3}}}$. To compare the relative performance of change point models and AI models, using true positive (TP), true negative (TN), false positive (FP), and false negative (FN), we have calculated; *accuracy=(TP+TN)/(TP+TN+FP+FN)*, *precision=TP/(TP+FP)*, and *detection=TP/(TP+FN)*. We have estimated the macro average values of precision and recall, which shows the average recall or precision values over the total number of classes. Area Under the Receiver Operating Characteristics (AUROC) is another performance measure that measures the model's aggregated performance to separate data points between binary classes (i.e., false information attack and no attack). AUROC is the area under the Receiver Operating Characteristics (ROC) curve, where the ROC means how good the model is to differentiate the true positive rate and false positive rate. Higher values of AUROC mean a better model. Figure 6 shows the performance of both change point and AI models, and the accuracy values are shown in bold font at the top of the bars in Figure 6. Based on the data used, BOCPD can detect false information attack with 99.9% accuracy without being sensitive to the hazard function, which is set at a constant rate of 0.10 per time interval. The CUSUM and EM models are also able to achieve good accuracy. Among the ML models, XGB model performed the best with 93.2% accuracy, followed by RF models achieving 89.7% accuracy. Although it requires the highest computational time to get trained, the NN model shows only 80% accuracy.

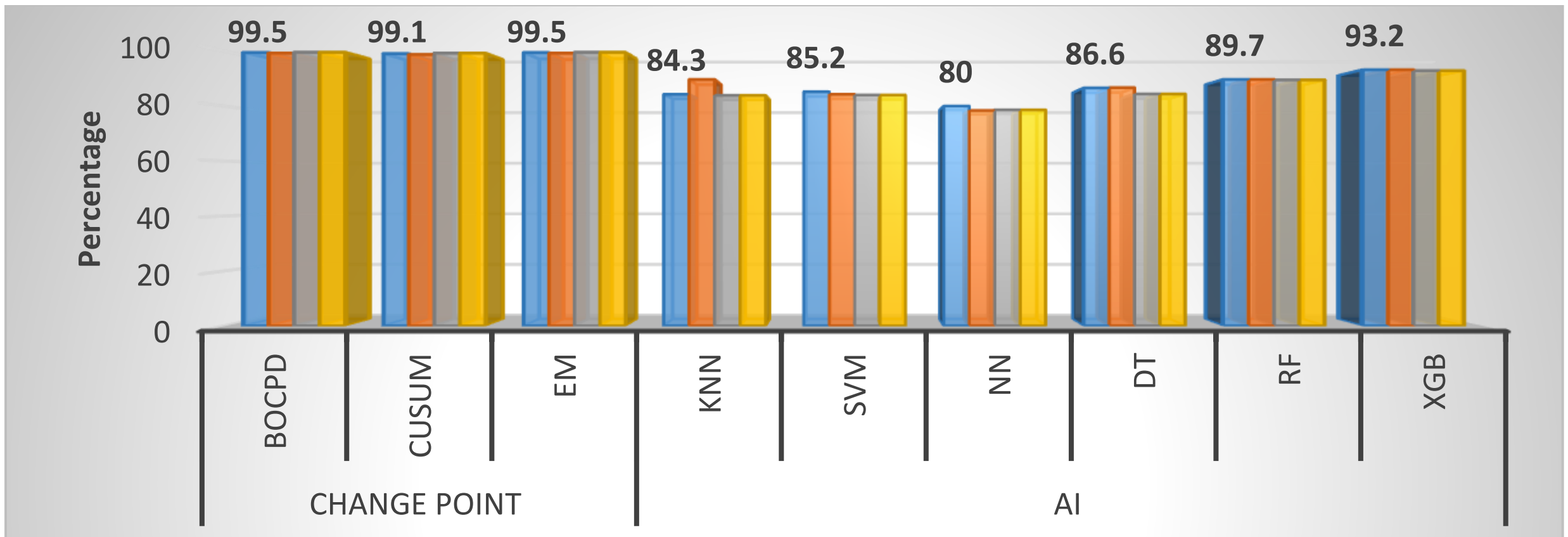

| | Change Point | | | AI | | | | | |
|---|---|---|---|---|---|---|---|---|---|
| | BOCPD | CUSUM | EM | KNN | SVM | NN | DT | RF | XGB |
| Accuracy (%) | 99.5 | 99.1 | 99.5 | 84.3 | 85.2 | 80 | 86.6 | 89.7 | 93.2 |
| Precision (macro) (%) | 99.3 | 98.8 | 99.3 | 89.7 | 84.3 | 78.4 | 86.7 | 89.7 | 93.2 |
| Recall (macro) (%) | 99.6 | 99.3 | 99.6 | 83.9 | 84 | 78.6 | 84.4 | 89.5 | 92.9 |
| AUROC (%) | 99.6 | 99.3 | 99.6 | 83.9 | 84 | 78.6 | 84.4 | 89.5 | 92.9 |

**Figure 6 Performance summary of the cyberattack detection models**

Overall, change point models perform much better in terms of all performance measures, which are accuracy, precision, recall, and AUROC. The reason behind this is that the change point models are more sensitive to the fluctuation of the CV BSM data, thus they can better detect the false information attack.

## CONCLUSIONS

Real-time false information cyberattack detection in connected vehicles is challenging as an attacker can dynamically attack any BSM data to inject malicious information that can lead to erroneous safety and operational impacts. To address such issues, change point models, or artificial intelligence models can be used for infrastructure edge-based false information cyberattack detection. Change point models benefit from no training requirement while still achieving accurate cyberattack detection accuracy compared with state-of-the-art AI models. By evaluating both change point and AI models, we have addressed the knowledge gap in the existing literature where the comparison between these models in cyberattack detection is missing. All change point models are able to detect false information attack with >99% accuracy in real-time. Although all AI models can also detect false information attacks in real-time, the extreme gradient boosting model has achieved the highest accuracy (93.2%), whereas the neural network model's accuracy is 80%.

Future research should be conducted considering multiple types cyberattack, such as denial of service and impersonation attacks, along with false information cyberattack on connected vehicles. We should also explore combining change point and AI models in cyberattack detection in connected vehicles Also, Change Point and AI models' performances for cyberattack could be evaluated in an environment where multiple cloud-based connected vehicle applications running simultaneously and providing different connected vehicle related services at the same time.

**AUTHOR CONTRIBUTIONS**

The authors confirm contribution to the paper as follows: study conception and design: MC, GC, SMK; data collection: SMK, GC; analysis and interpretation of results: MC, GC, SMK; draft manuscript preparation: MC, GC, SMK. All authors reviewed the results and approved the final version of the manuscript.